\documentclass{JHEP3}
\usepackage{amsmath}

\def\({\left(}
\def\){\right)}
\def\[{\left[}
\def\]{\right]}
\def\<{\langle}
\def\>{\rangle}
\def\ap{\alpha}
\def\bt{\beta}
\def\de{\delta}
\def\e{\varepsilon}
\def\la{\lambda}
\def\tla{\tilde{\lambda}}
\def\Sg{\Sigma}
\def\sg{\sigma}
\def\Th{\Theta}
\def\th{\theta}

\newcommand{\be}{\begin{equation}}
\newcommand{\ee}{\end{equation}}
\newcommand{\bal}{\begin{aligned}}
\newcommand{\eal}{\end{aligned}}
\newcommand{\barr}{\begin{array}}
\newcommand{\earr}{\end{array}}

\newcommand{\labell}[1]{\label{#1}}

\title{Note on Identities Inspired by New Soft Theorems}

\author{Junjie Rao$^{a}$, Bo Feng$^{ab}$\footnote{The
unconventional order is to let authors get proper
recognition of contributions under the outdated practice in China.}
\footnote{Emails: raojunjie@zju.edu.cn, b.feng@cms.zju.edu.cn} \\
{$^a$\small Zhejiang Institute of Modern Physics, Zhejiang University, Hangzhou, 310027, P. R. China \\
$^b$\small Center of Mathematical Science, Zhejiang University, Hangzhou, 310027, P. R. China \\}}

\abstract{The new soft theorems, for both gravity and gauge amplitudes, have inspired a number of works,
including the discovery of new identities related to amplitudes.
In this note, we present the proof and discussion for two sets of identities.
The first set includes an identity involving the half-soft function which had been used in the soft theorem
for one-loop rational gravity amplitudes, and another simpler identity as its byproduct.
The second set includes two identities involving the KLT momentum kernel, as the consistency conditions
of the KLT relation plus soft theorems for both gravity and gauge amplitudes. We use the CHY formulation to prove
the first identity, and transform the second one into a convenient form for future discussion.}

\keywords{Amplitudes, Soft Theorem}

\begin{document}
\maketitle

\section{Introduction}
\labell{intro}

Scattering amplitudes often have an universal soft behavior when the momentum of one
external leg tends to zero. This soft limit can be traced back to the works \cite{Low:54, Low:58, Weinberg}.
Recently, a new soft theorem for tree level gravity amplitudes was studied in \cite{Cachazo:2014fwa}.
By using the on-shell recursion relation \cite{Britto:2004ap, Britto:2005fq} and imposing the holomorphic soft limit,
Cachazo and Strominger have proved that
\be
\bal
&M_n\(\la_n\to\e\la_n\)\\
=&\frac{1}{\e^3}\sum_{a=1}^{n-2}\frac{\<n-1,a\>^2[na]}{\<n-1,n\>^2\<na\>}
M_{n-1}\(\tla_{n-1}\to\tla_{n-1}+\e\frac{\<an\>}{\<a,n-1\>}\tla_n,\,\tla_1\to\tla_1
+\e\frac{\<n-1,n\>}{\<n-1,a\>}\tla_n\)+O(\e^0),
\eal
\ee
here for $M_n$ and $M_{n-1}$, the unmentioned external kinematic data are un-deformed
and we prefer to suppress them for conciseness.
Taylor expansion in $\e$ exhibits three singular terms in orders $\e^{-3}$, $\e^{-2}$ and $\e^{-1}$,
while higher order terms in $\e$ will be mixed with the less interesting $O(\e^0)$ parts.

A similar relation for tree level Yang-Mills amplitudes using the on-shell recursion relation,
proved by Casali \cite{Casali:2014xpa}, takes the form
\be
\bal
&A_n\(\la_n\to\e\la_n\)\\
=&\frac{1}{\e^2}\frac{\<n-1,1\>}{\<n-1,n\>\<n1\>}
A_{n-1}\(\tla_{n-1}\to\tla_{n-1}+\e\frac{\<1n\>}{\<1,n-1\>}\tla_n,\,\tla_1\to\tla_1
+\e\frac{\<n-1,n\>}{\<n-1,1\>}\tla_n\)+O(\e^0),
\eal
\ee
where two singular terms in orders $\e^{-2}$ and $\e^{-1}$ appear
after Taylor expansion. The mixing between higher order terms from the
deformed $A_{n-1}$ and $O(\e^0)$ parts also persists to this case.

Based on this new discovery, many related studies have been done.
In \cite{Schwab:2014xua, Afkhami-Jeddi:2014fia, Zlotnikov:2014sva, Kalousios:2014uva, Schwab:2014fia,
Bianchi:2014gla, Schwab:2014sla, Chen:2014xoa, DiVecchia:2015oba, Bianchi:2015yta, Bork:2015fla, Chin:2015qza,
Campiglia:2015kxa, Bianchi:2015lnw}, the soft theorem has been generalized to arbitrary dimensions and other
theories or categories: string theory, ABJM theory, theories with fermions or massive particles, and form factors.
In \cite{Larkoski:2014hta, Adamo:2014yya, Geyer:2014lca, Kapec:2014opa, Broedel:2014fsa, Bern:2014vva, White:2014qia,
He:2014cra, Lysov:2014csa, Campiglia:2014yka, Liu:2014vva, Rao:2014zaa, Kapec:2014zla, Vera:2014tda, Mohd:2014oja,
Campiglia:2015yka, Pasterski:2015tva, Lipstein:2015rxa, Kapec:2015ena, Avery:2015gxa, Dumitrescu:2015fej},
the theorem has been understood from various perspectives,
especially those of symmetries and invariance.
In \cite{He:2014bga, Bianchi:2014gla, Bern:2014oka, Cachazo:2014dia, Brandhuber:2015vhm,
Bonocore:2014wua, Bonocore:2015esa},
its generalization to loop level has been discussed. In \cite{Chen:2014cuc, Cachazo:2015ksa, Klose:2015xoa,
Volovich:2015yoa, Du:2015esa, Georgiou:2015jfa, DiVecchia:2015bfa, Low:2015ogb}, the relevant double (or multiple)
soft theorem has also been discussed.

Among these studies, we have met two sets of identities which have not been proved so far.
We will present the proof in this note.

One identity of the first set was mentioned in \cite{He:2014bga}, which explored loop correction to the soft theorem.
It involves the so-called half-soft function $h$ (first defined in \cite{Bern:1998sv}
and reinterpreted in \cite{Feng:2012sy}), which appears naturally for all-plus one-loop gravity amplitude.
Its general proof was not given in \cite{He:2014bga}, but explicit checks up to 12 points had been done.
The identity reads
\be
\sum_{b\neq n}\<bn\>^2\sum_{M,N}h(b,n,M)h(b,n,N)\<b|K_M|n]\<n|K_N|b]^3=0, \labell{eq-1}
\ee
where $M,N$ are two nonempty partition sets of the $(n-2)$ particles other than $b$ and $n$, and $K_M$ and $K_N$
are the corresponding total momenta. During the proof, we had also discovered another simpler identity,
which can serve as its logical preliminary. It reads
\be
\frac{[1n]}{\<1n\>}\frac{|\psi_{N\cup M}|^w_w}{\<w1\>\<wn\>}
+\sum_N\<1|K_N|n]\<n|K_N|1]\frac{|\psi_N|^x_x}{\<x1\>\<xn\>}\frac{|\psi_M|^y_y}{\<y1\>\<yn\>}=0, \labell{eq-4}
\ee
where the $\psi$ matrix is related to $h$, and other symbols above will be explained shortly.

The second set of identities was conjectured in \cite{Du:2014eca}, which is a consequence of consistency conditions
between the soft theorems for gravity and gauge amplitudes, under the well-known KLT relation \cite{KLT:86}.
It involves the KLT momentum kernel \cite{Bern:1998sv, BjerrumBohr:2010ta, BjerrumBohr:2010zb, BjerrumBohr:2010yc},
and the transformation matrices ($D$ and $C$ below) between BCJ basis of gauge amplitudes \cite{Bern:2008qj}.
These two identities are
\be
\sum_{\ap_{t'},\bt_{t'}\in S_{n-3}}D[t,\ap_t,n-1,n|t',\ap_{t'},n-1,n]S[\ap_{t'}|\bt_{t'}]_{p_{n-1}}
C[t',n-1,\bt_{t'},n|t,n-1,\bt_t,n]=S[\ap_t|\bt_t]_{p_{n-1}}, \labell{eq-19}
\ee
\be
\sum_{t'=1}^{n-2}\sum_{\ap_{t'},\bt_{t'}\in S_{n-3}}D[t,\ap_t,n-1,n|t',\ap_{t'},n-1,n]S[\ap_{t'}|\bt_{t'}]_{p_{n-1}}
\cdot J_{t'}\(C[t',n-1,\bt_{t'},n|t,n-1,\bt_t,n]\)=0, \labell{eq-20}
\ee
where $S[\ap_t|\bt_t]_{p_{n-1}}$ is the KLT momentum kernel of pivot $p_{n-1}$,
and $J_{t'}\equiv J_{t',\dot{\ap}\dot{\bt}}$ is the anti-holomorphic angular momentum operator.
We will use the CHY formulation \cite{Cachazo:2013gna, Cachazo:2013hca, Cachazo:2013iea} to prove the first identity
and discuss the second one.

This note is organized as follows. In section \ref{sec2}, we prove
identity \eqref{eq-1} of the half-soft function, and also the byproduct identity \eqref{eq-4}.
In section \ref{sec3}, we prove identity \eqref{eq-19} of the KLT momentum kernel by
using the CHY formulation, while we transform identity \eqref{eq-20} into a convenient form
for possible future attempts and end with some discussion.

\section{Two Identities of the Half-soft Function}
\labell{sec2}

In this section we will prove \eqref{eq-1} and \eqref{eq-4}, first let's set up a bit convenient facilitation.
For reader's reference, we write \eqref{eq-1} again below
\be
\sum_{b\neq n}\<bn\>^2\sum_{M,N}h(b,n,M)h(b,n,N)\<b|K_M|n]\<n|K_N|b]^3=0,
\ee
where $M,N$ are two non-overlapping nonempty sets satisfying $M\cup N=\{1,\ldots,n-1\}\setminus b$,
and momentum conservation enforces $k_b+k_n+K_M+K_N=0$.
The half-soft function $h$ above is defined as \cite{Feng:2012sy}
\be
h(b,n,N)=\frac{1}{\prod^N_i\<ib\>^2\<in\>^2}|\Psi|^r_r=\frac{1}{\prod^N_i\<ib\>^2\<in\>^2}||\Psi||,
\ee
where $|\Psi|^r_r$ denotes the determinant of matrix $\Psi$ after deleting its $r$-th row and
$r$-th column, and $||\Psi||$ indicates this quantity is \textit{independent} of the choice $r\in N$.
If there is only one row and one column, the determinant is 1 after deletion.
The matrix $\Psi$ is defined as
\be
\Psi_{ij}(b,n)=-\frac{[ij]}{\<ij\>}\<ib\>\<in\>\<jb\>\<jn\>~\textrm{for}~i\neq j,~
\Psi_{ii}=\sum^N_{j\neq i}\Psi_{ij},
\ee
where $b$ and $n$ serve as auxiliary spinors. The sum of each row is zero, so $\Psi$ is degenerate.
Observe that the summand in \eqref{eq-1} has even power of $K_M$ and $K_N$, by momentum conservation
this sum is symmetric between $M$ and $N$, then we can replace $K_M$ by $-K_N$ and rewrite \eqref{eq-1} as
\be
\sum_{b\neq n}\<bn\>^2\sum_Nh(b,n,N)h(b,n,M)\<b|N|n]\<n|N|b]^3=0, \labell{eq-2}
\ee
for brevity $N$ stands for $K_N$ in spinorial products (and later $N$ also represents the number of elements
in the set $N$, depending on the context).

To simplify the proof, we define the matrix $\psi$ as
\be
\psi_{ij}(b,n)=-\frac{[ij]}{\<ij\>}\<jb\>\<jn\>~\textrm{for}~i\neq j,~
\psi_{ii}=\sum^N_{j\neq i}\psi_{ij},
\ee
where the common factor $\<ib\>\<in\>$ of the $i$-th row in $\Psi$ has been stripped off. One can easily verify that
\be
h(b,n,N)=\frac{1}{\prod^N_i\<ib\>^2\<in\>^2}||\Psi||
=\frac{1}{\prod^N_i\<ib\>\<in\>}\frac{|\psi_N|^x_x}{\<xb\>\<xn\>},
\ee
where $N$ has been added to $\psi$ to label the corresponding set, note that
$|\psi_N|^x_x/\<xb\>\<xn\>$ is \textit{independent} of the choice $x\in N$. Then we have
\be
h(b,n,N)h(b,n,M)=\frac{1}{\prod^{N+M}_i\<ib\>\<in\>}
\frac{|\psi_N|^x_x}{\<xb\>\<xn\>}\frac{|\psi_M|^y_y}{\<yb\>\<yn\>}
=\frac{\<bn\>}{\prod_{i\neq n}\<in\>\prod^{N+M}_i\<ib\>}
\frac{|\psi_N|^x_x}{\<xb\>\<xn\>}\frac{|\psi_M|^y_y}{\<yb\>\<yn\>},
\ee
where $\prod_{i\neq n}\<in\>$ is a common factor independent of $b$ so it can be dropped,
hence \eqref{eq-2} becomes
\be
\sum_{b\neq n}\frac{\<bn\>^3}{\prod_{i\neq b,n}\<ib\>}\sum_N\<b|N|n]\<n|N|b]^3
\frac{|\psi_N|^x_x}{\<xb\>\<xn\>}\frac{|\psi_M|^y_y}{\<yb\>\<yn\>}=0. \labell{eq-3}
\ee
%

\subsection{A simpler byproduct identity}

In the proof of \eqref{eq-3}, we happened to discover \eqref{eq-4}. For reader's reference, it is given below
\be
\frac{[1n]}{\<1n\>}\frac{|\psi_{N\cup M}|^w_w}{\<w1\>\<wn\>}
+\sum_N\<1|N|n]\<n|N|1]\frac{|\psi_N|^x_x}{\<x1\>\<xn\>}\frac{|\psi_M|^y_y}{\<y1\>\<yn\>}=0,
\ee
where $N,M$ are two non-overlapping nonempty sets satisfying $N\cup M=\{2,\ldots,n-1\}$,
and the auxiliary spinors are 1 and $n$. Also note that
$w\in N\cup M$, $x\in N$, $y\in M$ and it is free to switch the choices $w,x,y$ within each set.
Since this is mandatory for \eqref{eq-3} to hold, we will prove it first
as the tricks used here are analogous to those for \eqref{eq-3}.

Now we will adopt the BCFW deformation and reduce it into an identity of the same form,
but with one particle removed, in other words, we will perform an inductive proof. Before induction,
the identity is confirmed analytically at lower points for $n=4,5,6$.
For later convenience, we multiply it by a non-zero factor, yields
\be
\frac{1}{\prod_{i\neq 1,n}\<i1\>}\(\frac{[1n]}{\<1n\>}\frac{|\psi_{N\cup M}|^w_w}{\<w1\>\<wn\>}
+\sum_N\<1|N|n]\<n|N|1]\frac{|\psi_N|^x_x}{\<x1\>\<xn\>}\frac{|\psi_M|^y_y}{\<y1\>\<yn\>}\)=0, \labell{eq-5}
\ee
which is of course equivalent to \eqref{eq-4}. But now there are two advantages: The large $z$ behavior of its LHS
is improved, and it has the desired simple pole for residue evaluation, as we will soon see.

For generic $n$, consider BCFW deformation $\<1|n]$ and a particular pole $\<21\>$.
Note that particles 1 and $n$ are special while the rest $(n-2)$ ones are symmetric,
so it is sufficient to consider the residue of $\<21\>$ only,
as all $\<i1\>$'s with $i\in\{2,\ldots,n-1\}$ behave similarly. At $\<2\hat{1}\>=0$, we have
\be
|\hat{1}\>=|1\>-|n\>\frac{\<12\>}{\<n2\>}=|2\>\frac{\<1n\>}{\<2n\>},~
|\hat{n}]=|n]+|1]\frac{\<12\>}{\<n2\>}, \labell{eq-8}
\ee
and
\be
|\hat{1}\>[1|+|2\>[2|\equiv|2\>[\hat{2}|,~[\hat{2}|=[2|+[1|\frac{\<1n\>}{\<2n\>},
\ee
by which we mean to combine the momenta of particle $\hat{1}$ and 2 into that of particle $\hat{2}$,
or more physically, particles $\hat{1}$ and 2 merge into particle $\hat{2}$.
Including the deformed particle $\hat{n}$, the set $\{1,2,\ldots,n\}$ now
shrinks into $\{\hat{2},\ldots,\hat{n}\}$ while momentum conservation still holds, as what induction requires.

To locate pole $\<21\>$ in \eqref{eq-5}, we immediately find one in the overall factor. Naively, there might be
another one under $|\psi_N|^x_x$ if we take $x=2$, for example. However, the expansion
of $|\psi_N|^x_x$ in terms of $\<21\>$ will cancel this pole. In other words, $|\psi_N|^x_x/\<x1\>\<xn\>$
is a polynomial of $\<21\>$ (one may also choose $x\neq2$ to invalidate this pole),
that's why the overall factor is mandatory.

The next step is to analyze the large $z$ behavior of the LHS in \eqref{eq-5}
before evaluating its residues at finite locations. To clarify the analysis, we further separate
the second term in the parenthesis, and from now on we redefine
$N,M$ to exclude particle 2 from them while $N',M'$ denote the original sets.
Depending on whether $N'$ or $M'$ contains particle 2,
the set $\{2,\ldots,n-1\}$ has three types of splitting:
$\{\{2\}\cup N,M\}$, $\{N,\{2\}\cup M\}$ and $\{\{2\},N\cup M\}$, where $N\cup M=\{3,\ldots,n-1\}$.
So the second term becomes
\be
\bal
&\sum_{N'}\<1|N'|n]\<n|N'|1]\frac{|\psi_{N'}|^x_x}{\<x1\>\<xn\>}\frac{|\psi_{M'}|^y_y}{\<y1\>\<yn\>}\\
=&\sum_{\{2\}\cup N,M}(\<1|N|n]+\<12\>[2n])(\<n|N|1]+\<n2\>[21])
\frac{|\psi_{N\cup\{2\}}|^2_2}{\<21\>\<2n\>}\frac{|\psi_M|^y_y}{\<y1\>\<yn\>}\\
&+\sum_{N,\{2\}\cup M}\<1|N|n]\<n|N|1]\frac{|\psi_N|^x_x}{\<x1\>\<xn\>}\frac{|\psi_{M\cup\{2\}}|^2_2}{\<21\>\<2n\>}
+[n2][12]\frac{|\psi_{N\cup M}|^w_w}{\<w1\>\<wn\>}. \labell{eq-6}
\eal
\ee
Also, the first term in \eqref{eq-5} can be written as
\be
\frac{[1n]}{\<1n\>}\frac{|\psi_{N'\cup M'}|^w_w}{\<w1\>\<wn\>}
=\frac{[1n]}{\<1n\>}\frac{|\psi_{N\cup M\cup\{2\}}|^2_2}{\<21\>\<2n\>}. \labell{eq-7}
\ee
Since the three $\psi$'s in \eqref{eq-7} and the first and second terms of \eqref{eq-6} contain particle 2,
we can choose to delete its corresponding row and column.
Large $z$ power counting shows that all four terms in \eqref{eq-6} and
\eqref{eq-7} behave as $z^{N+M-1}=z^{n-4}$ under $\<1|n]$, but the overall factor in the front of \eqref{eq-5}
behaves as $z^{-(n-2)}$, which renders the entire expression as $z^{-2}$, so there is no boundary contribution.
Therefore, via contour integration, the LHS of \eqref{eq-5} (denoted $I$ below) can be expressed as
\be
\oint_{z=0}\frac{dz}{z}\,I(z)=-\oint_{\<2\hat{1}\>=0}\frac{dz}{z}\,I(z)-\ldots
-\oint_{\<n-1,\hat{1}\>=0}\frac{dz}{z}\,I(z), \labell{eq-23}
\ee
if the residue at $\<2\hat{1}\>=0$ vanishes, by the symmetry among particles $\{2,\ldots,n-1\}$ the entire un-deformed
expression must also vanish.
Note the contribution from the overall factor in \eqref{eq-5} is universal, so it can be dropped.
At $\<2\hat{1}\>=0$, after some algebra, the residue evaluation gives
\be
\frac{|\psi_{N\cup\{2\}}|^2_2}{\<21\>\<2n\>}\to\(\frac{\<1n\>}{\<2n\>}\)^{N-1}(-\<n|N|2])
\frac{|\psi_N|^x_x}{\<x2\>\<xn\>}, \labell{eq-9}
\ee
recall that $\<21\>$ above is not a pole, while the real pole comes from the overall factor.
Here $|\hat{1}\>$ is replaced by $|2\>$ up to a factor, after recalling \eqref{eq-8}. By expanding the determinant
to the first order of $\<21\>$, then using the independence of choice $x$ to switch the deleted row and column
for each term, we can collect a factor $(-\<n|N|2])$ as above. The similar (and simpler) story happens for
\be
\frac{|\psi_M|^y_y}{\<y1\>\<yn\>}\to\(\frac{\<1n\>}{\<2n\>}\)^{M-2}
\frac{|\psi_M|^y_y}{\<y2\>\<yn\>}. \labell{eq-10}
\ee
Plugging them back, up to a factor $(\<1n\>/\<2n\>)^{N+M-2}$, the sum of \eqref{eq-6} and \eqref{eq-7} becomes
\be
\bal
&(-\<n|N+M|2])\frac{|\psi_{N\cup M}|^w_w}{\<w2\>\<wn\>}\frac{[1n]}{\<2n\>}+\sum_N\<2|N|\hat{n}]
(\<n|N|1]+\<n2\>[21])(-\<n|N|2])\frac{|\psi_N|^x_x}{\<x2\>\<xn\>}\frac{|\psi_M|^y_y}{\<y2\>\<yn\>}\\
&+\sum_N\<2|N|\hat{n}]\<n|N|1](-\<n|M|2])\frac{|\psi_N|^x_x}{\<x2\>\<xn\>}\frac{|\psi_M|^y_y}{\<y2\>\<yn\>}
+[\hat{n}2][12]\frac{|\psi_{N\cup M}|^w_w}{\<w2\>\<wn\>}.
\eal
\ee
By momentum conservation, up to a factor $[12]$, it can be simplified into
\be
\bal
&\(\frac{\<1n\>}{\<2n\>}[n1]+[\hat{n}2]\)\frac{|\psi_{N\cup M}|^w_w}{\<w2\>\<wn\>}
+\<n2\>\sum_N\<2|N|\hat{n}]\(\<n|N|2]+\<n|N|1]\frac{\<1n\>}{\<2n\>}\)
\frac{|\psi_N|^x_x}{\<x2\>\<xn\>}\frac{|\psi_M|^y_y}{\<y2\>\<yn\>}\\
=\,&\<n2\>\(\frac{[\hat{2}\hat{n}]}{\<2n\>}\frac{|\psi_{N\cup M}|^w_w}{\<w2\>\<wn\>}
+\sum_N\<2|N|\hat{n}]\<n|N|\hat{2}]\frac{|\psi_N|^x_x}{\<x2\>\<xn\>}\frac{|\psi_M|^y_y}{\<y2\>\<yn\>}\)=0,
\eal
\ee
after assuming the identity of $(n-1)$ particles holds. This finishes the inductive proof of \eqref{eq-4}.

\subsection{Proof of the first identity}

Now we move to prove \eqref{eq-3} by applying the similar pack of tricks: to consider deformation $\<1|n]$
acting on its LHS, and the pole $\<21\>$. First, we separate the expression
into three parts corresponding to $b=1$, $b=2$ and $b=3,\ldots,n-1$, namely
\be
\bal
\sum_{b\neq n}\frac{\<bn\>^3}{\prod_{i\neq b,n}\<ib\>}\sum_N\<b|N|n]\<n|N|b]^3
\frac{|\psi_N|^x_x}{\<xb\>\<xn\>}\frac{|\psi_M|^y_y}{\<yb\>\<yn\>}=I_1+I_2+I_{b\neq1,2,n}.
\eal
\ee
Similarly, we now redefine $N$ and $M$ to exclude particles 2 and 1, with respect to $I_1$ and $I_2$.
For $I_1$, the set $\{2,\ldots,n-1\}$ has
three types of splitting: $\{\{2\}\cup N,M\}$, $\{N,\{2\}\cup M\}$ and $\{\{2\},N\cup M\}$,
where $N\cup M=\{3,\ldots,n-1\}$. For $I_2$, we have $\{\{1\}\cup N,M\}$, $\{N,\{1\}\cup M\}$ and $\{\{1\},N\cup M\}$.
For $I_{b\neq1,2,n}$, there are four types: $\{\{1,2\}\cup N_b,M_b\}$, $\{N_b,\{1,2\}\cup M_b\}$,
$\{\{1\}\cup N_b,\{2\}\cup M_b\}$ and $\{\{2\}\cup N_b,\{1\}\cup M_b\}$,
where $N_b\cup M_b=\{3,\ldots,n-1\}\setminus b$, but the last two will not contribute to the residue of $\<21\>$
and hence the corresponding terms are neglected, which will be explained shortly.

According to the splittings above, we can write
\be
\bal
\frac{I_1}{\<1n\>^3}=\,&\frac{1}{\prod_{i\neq1,2,n}\<i1\>\<21\>}\sum_N\(\<1|N+2|n]\<n|N+2|1]^3
\frac{|\psi_{N\cup\{2\}}|^2_2}{\<21\>\<2n\>}\frac{|\psi_M|^y_y}{\<y1\>\<yn\>}+\<1|N|n]\<n|N|1]^3
\frac{|\psi_N|^x_x}{\<x1\>\<xn\>}\frac{|\psi_{M\cup\{2\}}|^2_2}{\<21\>\<2n\>}\)\\
&+\frac{1}{\prod_{i\neq1,2,n}\<i1\>\<21\>}\<1|2|n]\<n|2|1]^3\frac{1}{\<21\>\<2n\>}
\frac{|\psi_{N\cup M}|^w_w}{\<w1\>\<wn\>},
\eal
\ee
\be
\bal
\frac{I_2}{\<2n\>^3}=\,&\frac{1}{\prod_{i\neq1,2,n}\<i2\>\<12\>}\sum_N\(\<2|N+1|n]\<n|N+1|2]^3
\frac{|\psi_{N\cup\{1\}}|^1_1}{\<12\>\<1n\>}\frac{|\psi_M|^y_y}{\<y2\>\<yn\>}+\<2|N|n]\<n|N|2]^3
\frac{|\psi_N|^x_x}{\<x2\>\<xn\>}\frac{|\psi_{M\cup\{1\}}|^1_1}{\<12\>\<1n\>}\)\\
&+\frac{1}{\prod_{i\neq1,2,n}\<i2\>\<12\>}\<2|1|n]\<n|1|2]^3\frac{1}{\<12\>\<1n\>}
\frac{|\psi_{N\cup M}|^w_w}{\<w2\>\<wn\>},
\eal
\ee
\be
\bal
I_{b\neq1,2,n}=&\sum_{b\neq1,2,n}\frac{\<bn\>^3}{\prod_{i\neq1,b,n}\<ib\>\<1b\>}\\
&\times\sum_{N_b}\(\<b|N_b+1+2|n]\<n|N_b+1+2|b]^3
\frac{|\psi_{N_b\cup\{1,2\}}|^1_1}{\<1b\>\<1n\>}\frac{|\psi_{M_b}|^y_y}{\<yb\>\<yn\>}
+\<b|N_b|n]\<n|N_b|b]^3\frac{|\psi_{N_b}|^x_x}{\<xb\>\<xn\>}\frac{|\psi_{M_b\cup\{1,2\}}|^1_1}{\<1b\>\<1n\>}\)\\
&+(\textrm{two neglected terms}).
\labell{eq-12}
\eal
\ee
For $I_{b\neq1,2,n}$ one can verify that, only terms for which 1 and 2 are in the same splitting set,
have pole $\<21\>$ and hence contribute to the residue, which explains why we only need the first two terms.
Moreover, $N_b$ in $\{\{1,2\}\cup N_b,M_b\}$ can be empty (similarly for $M_b$).
While for $I_1$, $N$ in $\{\{2\}\cup N,M\}$ cannot be empty,
otherwise such a splitting belongs to type $\{\{2\},N\cup M\}$ (similarly for $I_2$).

After the separation, we now analyze the large $z$ behavior.
Under $\<1|n]$, large $z$ power counting shows that $I_1\sim z^{-2}$, $I_2\sim z^{-1}$
and $I_{b\neq1,2,n}\sim z^{-1}$, so there is no boundary contribution.
Then we can repeat the contour integration \eqref{eq-23}.
Again, thanks to the symmetry among particles $\{2,\ldots,n-1\}$,
it is sufficient to consider the residue of $\<21\>$ only.

Recalling \eqref{eq-9} and \eqref{eq-10}, at $\<2\hat{1}\>=0$ the residue evaluation gives
\be
\bal
\frac{\<21\>}{\<1n\>^3}I_1\to\,&\frac{\<2n\>^3}{\<1n\>^3}\frac{1}{\prod_{i\neq1,2,n}\<i2\>}\\
&\times\sum_N\[\<\hat{1}|N+2|\hat{n}]\<n|N+2|1]^3(-\<n|N|2])
+\<\hat{1}|N|\hat{n}]\<n|N|1]^3(-\<n|M|2])\]\frac{|\psi_N|^x_x}{\<x2\>\<xn\>}\frac{|\psi_M|^y_y}{\<y2\>\<yn\>}\\
&-\frac{\<2n\>^4}{\<1n\>^2}\frac{[12]^3[2\hat{n}]}{\prod_{i\neq1,2,n}\<i2\>}\frac{|\psi_{N\cup M}|^w_w}{\<w2\>\<wn\>},
\eal
\ee
or after a bit simplification,
\be
\bal
\frac{\<21\>\<1n\>}{\<2n\>^2}I_1\to\,&\frac{\<1n\>^2}{\prod_{i\neq1,2,n}\<i2\>}\sum_N\<2|N|\hat{n}]
\(-\<n|N|2]\<n|N+2|1]^3+\<n|N+1|2]\<n|N|1]^3\)\frac{|\psi_N|^x_x}{\<x2\>\<xn\>}\frac{|\psi_M|^y_y}{\<y2\>\<yn\>}\\
&-\<1n\>^2\<2n\>^2\frac{[12]^3[2\hat{n}]}{\prod_{i\neq1,2,n}\<i2\>}\frac{|\psi_{N\cup M}|^w_w}{\<w2\>\<wn\>}.
\eal
\ee
Similarly for $I_2$,
\be
\bal
\frac{\<21\>\<1n\>}{\<2n\>^2}I_2\to\,&\frac{\<2n\>^2}{\prod_{i\neq1,2,n}\<i2\>}\sum_N\<2|N|\hat{n}]
\(\<n|N|1]\<n|N+1|2]^3-\<n|N+2|1]\<n|N|2]^3\)\frac{|\psi_N|^x_x}{\<x2\>\<xn\>}\frac{|\psi_M|^y_y}{\<y2\>\<yn\>}\\
&-\<1n\>^3\<2n\>\frac{[12]^3[1n]}{\prod_{i\neq1,2,n}\<i2\>}\frac{|\psi_{N\cup M}|^w_w}{\<w2\>\<wn\>}.
\eal
\ee
Combining $I_1$ and $I_2$, we find
\be
\bal
\frac{\<21\>\<1n\>}{\<2n\>^2}(I_1+I_2)
\to&-\frac{[12]\<2n\>^3}{\prod_{i\neq1,2,n}\<i2\>}\sum_N\<2|N|\hat{n}]\<n|N|\hat{2}]^3
\frac{|\psi_N|^x_x}{\<x2\>\<xn\>}\frac{|\psi_M|^y_y}{\<y2\>\<yn\>}\\
&-\frac{[12]^3\<1n\>^2\<2n\>^3}{\prod_{i\neq1,2,n}\<i2\>}\sum_N\<2|N|\hat{n}]\<n|N|\hat{2}]
\frac{|\psi_N|^x_x}{\<x2\>\<xn\>}\frac{|\psi_M|^y_y}{\<y2\>\<yn\>}
-\<1n\>^2\<2n\>^2\frac{[12]^3[\hat{2}\hat{n}]}{\prod_{i\neq1,2,n}\<i2\>}\frac{|\psi_{N\cup M}|^w_w}{\<w2\>\<wn\>},
\labell{eq-11}
\eal
\ee
after using the following identity
\be
\bal
&\<1n\>^2\(-\<n|N|2]\<n|N+2|1]^3+\<n|N+1|2]\<n|N|1]^3\)
+\<2n\>^2\(\<n|N|1]\<n|N+1|2]^3-\<n|N+2|1]\<n|N|2]^3\)\\
=&-[12](\<1n\>\<n|N|1]+\<2n\>\<n|N|2])^3-[12]^3\<1n\>^2\<2n\>^2(\<1n\>\<n|N|1]+\<2n\>\<n|N|2])\\
=&-[12]\<2n\>^3\<n|N|\hat{2}]^3-[12]^3\<1n\>^2\<2n\>^3\<n|N|\hat{2}].
\eal
\ee
Now note the second and third terms in \eqref{eq-11} can be regrouped as
\be
-\frac{[12]^3\<1n\>^2\<2n\>^3}{\prod_{i\neq1,2,n}\<i2\>}\(\sum_N\<2|N|\hat{n}]\<n|N|\hat{2}]
\frac{|\psi_N|^x_x}{\<x2\>\<xn\>}\frac{|\psi_M|^y_y}{\<y2\>\<yn\>}
+\frac{[\hat{2}\hat{n}]}{\<2n\>}\frac{|\psi_{N\cup M}|^w_w}{\<w2\>\<wn\>}\)=0,
\ee
which is exactly identity \eqref{eq-4} for the set $\{\hat{2},\ldots,\hat{n}\}$! Therefore we are left with
\be
\frac{\<21\>\<1n\>}{\<2n\>^2}(I_1+I_2)
\to\,-\frac{[12]\<2n\>^3}{\prod_{i\neq1,2,n}\<i2\>}\sum_N\<2|N|\hat{n}]\<n|N|\hat{2}]^3
\frac{|\psi_N|^x_x}{\<x2\>\<xn\>}\frac{|\psi_M|^y_y}{\<y2\>\<yn\>}. \labell{eq-13}
\ee
To settle this leftover, we look back to $I_{b\neq1,2,n}$ in \eqref{eq-12} and find
\be
\frac{|\psi_{N_b\cup\{1,2\}}|^1_1}{\<1b\>\<1n\>}\to\frac{[21]}{\<21\>}|\psi_{N_b\cup\{\hat{2}\}}|^{\hat{2}}_{\hat{2}}
=-\frac{[12]}{\<21\>}\<2b\>\<2n\>\frac{|\psi_{N_b\cup\{\hat{2}\}}|^x_x}{\<xb\>\<xn\>},
\ee
where again we have used the independence of choice $x$ to switch the deleted row and column. Now
\be
\bal
\frac{\<21\>\<1n\>}{\<2n\>^2}I_{b\neq1,2,n}\to&-[12]\sum_{b\neq1,2,n}\frac{\<bn\>^3}{\prod_{i\neq1,b,n}\<ib\>}\\
&\times\sum_{N_b}\(\<b|N_b+\hat{2}|\hat{n}]\<n|N_b+\hat{2}|b]^3
\frac{|\psi_{N_b\cup\{\hat{2}\}}|^x_x}{\<xb\>\<xn\>}\frac{|\psi_{M_b}|^y_y}{\<yb\>\<yn\>}
+\<b|N_b|\hat{n}]\<n|N_b|b]^3\frac{|\psi_{N_b}|^x_x}{\<xb\>\<xn\>}\frac{|\psi_{M_b\cup\{\hat{2}\}}|^y_y}{\<yb\>\<yn\>}\).
\labell{eq-14}
\eal
\ee
Summing \eqref{eq-13} and \eqref{eq-14}, we get
\be
\bal
&-\frac{\<21\>\<1n\>}{\<2n\>^2[12]}(I_1+I_2+I_{b\neq1,2,n})\\
\to&\,\frac{\<2n\>^3}{\prod_{i\neq1,2,n}\<i2\>}
\sum_N\<2|N|\hat{n}]\<n|N|\hat{2}]^3\frac{|\psi_N|^x_x}{\<x2\>\<xn\>}\frac{|\psi_M|^y_y}{\<y2\>\<yn\>}
+\sum_{b\neq1,2,n}\frac{\<bn\>^3}{\prod_{i\neq1,b,n}\<ib\>}
\sum_{N'_b}\<b|N'_b|\hat{n}]\<n|N'_b|b]^3\frac{|\psi_{N'_b}|^x_x}{\<xb\>\<xn\>}
\frac{|\psi_{M'_b}|^y_y}{\<yb\>\<yn\>}\\
=&\sum_{b\neq1,n}\frac{\<bn\>^3}{\prod_{i\neq1,b,n}\<ib\>}
\sum_{N'}\<b|N'|\hat{n}]\<n|N'|b]^3\frac{|\psi_{N'}|^x_x}{\<xb\>\<xn\>}
\frac{|\psi_{M'}|^y_y}{\<yb\>\<yn\>}=0,
&
\eal
\ee
which returns to the form of \eqref{eq-3} for the set $\{\hat{2},\ldots,\hat{n}\}$!
It vanishes after assuming the identity of $(n-1)$ particles (without particle 1) holds.
Similar to $N',M'$, here $N'_b,M'_b$ denote the sets including $\hat{2}$ but not $b$.
This finishes the inductive proof of \eqref{eq-1}.

\section{Two Identities of the KLT Momentum Kernel}
\labell{sec3}

In this section we will prove \eqref{eq-19} and \eqref{eq-20} as conjectured in \cite{Du:2014eca}.
To understand these relations, we must first define the transformation matrices $D$ and $C$ between
BCJ basis of gauge amplitudes via
\be
A_n(t,\ap_t,n-1,n)=\sum_{\ap_{t'}\in S_{n-3}}A_n(t',\ap_{t'},n-1,n)D[t',\ap_{t'},n-1,n|t,\ap_t,n-1,n], \labell{eq-17}
\ee
\be
\widetilde{A}_n(t,n-1,\bt_t,n)=\sum_{\bt_{t'}\in S_{n-3}}C[t,n-1,\bt_t,n|t',n-1,\bt_{t'},n]
\widetilde{A}_n(t',n-1,\bt_{t'},n), \labell{eq-18}
\ee
where $\ap_{t'}$ and $\bt_{t'}$ denote the permutations of $(n-3)$ particles other than $t'$, $(n-1)$ and $n$.
In a tensorial sense, $D$ and $C$ are the transformation matrices with respect to the summation of all
$(n-3)!$ permutations, which is defined as the inner product. For reader's reference,
we write \eqref{eq-19} and \eqref{eq-20} again below
\be
\sum_{\ap_{t'},\bt_{t'}\in S_{n-3}}D[t,\ap_t,n-1,n|t',\ap_{t'},n-1,n]S[\ap_{t'}|\bt_{t'}]_{p_{n-1}}
C[t',n-1,\bt_{t'},n|t,n-1,\bt_t,n]=S[\ap_t|\bt_t]_{p_{n-1}},
\ee
\be
\sum_{t'=1}^{n-2}\sum_{\ap_{t'},\bt_{t'}\in S_{n-3}}D[t,\ap_t,n-1,n|t',\ap_{t'},n-1,n]S[\ap_{t'}|\bt_{t'}]_{p_{n-1}}
\cdot J_{t'}\(C[t',n-1,\bt_{t'},n|t,n-1,\bt_t,n]\)=0,
\ee
where $S[\ap_t|\bt_t]_{p_{n-1}}$ is the KLT momentum kernel of pivot $p_{n-1}$,
and $J_{t'}\equiv J_{t',\dot{\ap}\dot{\bt}}$ is the anti-holomorphic angular momentum operator.
Here we follow the convention of $S$ in \cite{BjerrumBohr:2010ta, BjerrumBohr:2010zb, BjerrumBohr:2010yc}, namely
\be
S[\ap_1,\ldots,\ap_k|\bt_1,\ldots,\bt_k]_{p_{n-1}}
=\prod_{i=1}^k\(s_{\ap_i,n-1}+\sum_{j=i+1}^k\th(\ap_i,\ap_j)s_{\ap_i,\ap_j}\),
\ee
where $s_{ij}$ is each Mandelstam variable, and $\th(\ap_i,\ap_j)$ is zero when the pair $(\ap_i,\ap_j)$ has the
same ordering at both sets $\{\ap_1,\ldots,\ap_k\}$ and $\{\bt_1,\ldots,\bt_k\}$, and unity otherwise.

For the first identity, its physical interpretation is straightforward: If we regard the KLT momentum kernel $S$ as
the metric, it is simply the tensorial transformation rule for metric. In fact, such a tensorial formulation
had been established in \cite{Cachazo:2013gna, Cachazo:2013iea}
(known as the KLT orthogonality or the CHY formulation) and we will use it to formally prove the first identity
shortly. The second identity is however more intricate, as it roughly represents angular momentum conservation
in an entangled way. The CHY formulation can help transform it into a relation that may reveal very nontrivial
properties of scattering process, while to prove it directly is yet beyond our understanding.

\subsection{Proof of the first identity}

Before the proof, we must first rewrite gauge amplitudes in the CHY formulation \cite{Cachazo:2013iea}
which is based on the scattering equations \cite{Cachazo:2013hca}. It tells that
\be
A_n(t,\ap_t,n-1,n)=\sum_{i=1}^{(n-3)!}
\frac{1}{\det'(\Phi)(\sg^{(i)})}\Sg^{(i)}(t,\ap_t,n-1,n)\,\textrm{Pf}\,'\Psi(\sg^{(i)}), \labell{eq-15}
\ee
\be
\widetilde{A}_n(t,n-1,\bt_t,n)=\sum_{i=1}^{(n-3)!}\frac{1}{\det'(\Phi)(\sg^{(i)})}\Sg^{(i)}(t,n-1,\bt_t,n)
\,\textrm{Pf}\,'\Psi(\sg^{(i)}), \labell{eq-16}
\ee
where $\sg^{(i)}$ denotes the $i$-th solution to the scattering equations
\be
\sum_{b\neq a}\frac{s_{ab}}{\sg_{ab}}=0,
\ee
with $\sg_{ab}=\sg_a-\sg_b$, and there are $(n-3)!$ solutions in total.
The definitions of $\det'(\Phi)$ and $\textrm{Pf}\,'\Psi$, namely the reduced determinant of Jacobian $\Phi$
and the reduced Pffafian of antisymmetric matrix $\Psi$,
can be found in \cite{Cachazo:2013iea}. The object mainly concerns us is
\be
\Sg^{(i)}(\ap)\equiv\frac{1}{\sg^{(i)}_{\ap(1),\ap(2)}\ldots\sg^{(i)}_{\ap(n-1),\ap(n)}\sg^{(i)}_{\ap(n),\ap(1)}}.
\ee
On the other hand, the KLT relation gives
\be
\bal
(-)^{n+1}M_n(1,\ldots,n)=&\sum_{\ap_t,\bt_t\in S_{n-3}}
A_n(t,\ap_t,n-1,n)S[\ap_t|\bt_t]_{p_{n-1}}\widetilde{A}_n(t,n-1,\bt_t,n)\\
=&\sum_{i=1}^{(n-3)!}
\frac{\textrm{Pf}\,'\Psi(\sg^{(i)})\textrm{Pf}\,'\Psi(\sg^{(i)})}{\det'(\Phi)(\sg^{(i)})},
\eal
\ee
where the second line results from the CHY formulation. There is a subtle issue of the sign above,
due to the different conventions $M_n=-M_n^\textrm{CHY}$ and $S[\ap_t|\bt_t]=S^\textrm{CHY}[\bt_t|\ap_t^\textrm{T}]$.
Plugging \eqref{eq-15} and \eqref{eq-16} into this relation, yields
\be
\sum_{\ap_t,\bt_t\in S_{n-3}}\Sg^{(i)}(t,\ap_t,n-1,n)S[\ap_t|\bt_t]_{p_{n-1}}\Sg^{(j)}(t,n-1,\bt_t,n)
={\det}'(\Phi)(\sg^{(i)})\de_{ij},
\ee
or more compactly,
\be
G_{i\ap_t}S[\ap_t|\bt_t]\(H_{j\bt_t}\)^\textrm{T}=I_{(n-3)!\times(n-3)!},
\ee
which is the KLT orthogonality, if we define matrices
\be
G_{i\ap_t}\equiv\frac{\Sg^{(i)}(t,\ap_t,n-1,n)}{\sqrt{\det'(\Phi)(\sg^{(i)})}},~
H_{j\bt_t}\equiv\frac{\Sg^{(j)}(t,n-1,\bt_t,n)}{\sqrt{\det'(\Phi)(\sg^{(j)})}}.
\ee
From this matrix relation we immediately get
\be
S[\ap_t|\bt_t]=\(G_{i\ap_t}\)^{-1}\(\(H_{i\bt_t}\)^\textrm{T}\)^{-1}. \labell{eq-21}
\ee
Back to \eqref{eq-15} and \eqref{eq-16}, if we further define the row vector
\be
\Th_i\equiv\frac{\textrm{Pf}\,'\Psi(\sg^{(i)})}{\sqrt{\det'(\Phi)(\sg^{(i)})}},
\ee
then
\be
A_n(t,\ap_t,n-1,n)=\Th_iG_{i\ap_t},~~\widetilde{A}_n(t,n-1,\bt_t,n)=\Th_iH_{i\bt_t}.
\ee
Plugging them back into \eqref{eq-17} and \eqref{eq-18}, and assuming their independence of basis $\Th_i$, we get
\be
G_{i\ap_t}=G_{i\ap_{t'}}D[t',\ap_{t'},n-1,n|t,\ap_t,n-1,n],~~
H_{i\bt_t}=H_{i\bt_{t'}}\(C[t,n-1,\bt_t,n|t',n-1,\bt_{t'},n]\)^\textrm{T},
\ee
or equivalently,
\be
D[t',\ap_{t'},n-1,n|t,\ap_t,n-1,n]=\(G_{i\ap_{t'}}\)^{-1}G_{i\ap_t},~~
C[t,n-1,\bt_t,n|t',n-1,\bt_{t'},n]=(H_{i\bt_t})^\textrm{T}\((H_{i\bt_{t'}})^{-1}\)^\textrm{T}.
\ee
Finally we plug them back into the LHS of \eqref{eq-19} and interchange $t$ and $t'$,
together with \eqref{eq-21} we get
\be
\bal
&\sum_{\ap_{t'},\bt_{t'}\in S_{n-3}}D[t,\ap_t,n-1,n|t',\ap_{t'},n-1,n]S[\ap_{t'}|\bt_{t'}]_{p_{n-1}}
C[t',n-1,\bt_{t'},n|t,n-1,\bt_t,n]\\
=&\(G_{i\ap_t}\)^{-1}G_{i\ap_{t'}}\(G_{j\ap_{t'}}\)^{-1}\(\(H_{j\bt_{t'}}\)^\textrm{T}\)^{-1}
(H_{k\bt_{t'}})^\textrm{T}\((H_{k\bt_t})^{-1}\)^\textrm{T}\\
=&\(G_{i\ap_t}\)^{-1}\(\(H_{i\bt_t}\)^\textrm{T}\)^{-1}=S[\ap_t|\bt_t]_{p_{n-1}},
\eal
\ee
which is exactly the RHS of \eqref{eq-19}, hence the proof is finished.

\subsection{Discussion of the second identity}

Now we move to prove \eqref{eq-20}.
Equipped with the matrices defined in the previous subsection, the LHS of \eqref{eq-20}
can be simplified as
\be
\bal
&\sum_{t'=1}^{n-2}\(G_{i\ap_t}\)^{-1}G_{i\ap_{t'}}\(G_{j\ap_{t'}}\)^{-1}\(\(H_{j\bt_{t'}}\)^\textrm{T}\)^{-1}
\cdot J_{t'}\[(H_{k\bt_{t'}})^\textrm{T}\((H_{k\bt_t})^{-1}\)^\textrm{T}\]\\
=&\sum_{t'=1}^{n-2}\(G_{i\ap_t}\)^{-1}\(\(H_{i\bt_{t'}}\)^\textrm{T}\)^{-1}
\cdot J_{t'}\[(H_{j\bt_{t'}})^\textrm{T}\((H_{j\bt_t})^{-1}\)^\textrm{T}\]\\
=&\sum_{t'=1}^{n-2}\(G_{i\ap_t}\)^{-1}\[\(\(H_{i\bt_{t'}}\)^\textrm{T}\)^{-1}
(H_{j\bt_{t'}})^\textrm{T}\cdot J_{t'}\((H_{j\bt_t})^{-1}\)^\textrm{T}
+\(\(H_{i\bt_{t'}}\)^\textrm{T}\)^{-1}\cdot J_{t'}(H_{j\bt_{t'}})^\textrm{T}
\cdot \((H_{j\bt_t})^{-1}\)^\textrm{T}\]\\
=&\sum_{t'=1}^{n-2}\(G_{i\ap_t}\)^{-1}\[J_{t'}\((H_{i\bt_t})^{-1}\)^\textrm{T}\cdot(H_{j\bt_t})^\textrm{T}
+\(\(H_{i\bt_{t'}}\)^\textrm{T}\)^{-1}\cdot J_{t'}(H_{j\bt_{t'}})^\textrm{T}\]\((H_{j\bt_t})^{-1}\)^\textrm{T},
\eal
\ee
assuming the two matrices in the front and end of the last line are non-degenerate, we should prove
\be
\sum_{t'=1}^{n-2}\[J_{t'}\((H_{i\bt_t})^{-1}\)^\textrm{T}\cdot(H_{j\bt_t})^\textrm{T}
+\(\(H_{i\bt_{t'}}\)^\textrm{T}\)^{-1}\cdot J_{t'}(H_{j\bt_{t'}})^\textrm{T}\]=0.
\ee
For the first term above, the summation over $t'$ is trivial since the matrix product involves $t$ only,
so it is in fact
\be
\(\sum_{t'=1}^{n-2}J_{t'}\)\((H_{i\bt_t})^{-1}\)^\textrm{T}\cdot(H_{j\bt_t})^\textrm{T}=0,
\ee
due to angular momentum conservation, as the absence of $J_{n-1}$ and $J_n$ does not matter
since $\tla_{n-1}$ and $\tla_n$ have been solved by momentum conservation (see \cite{Du:2014eca} for more details).
Therefore we are left with
\be
\sum_{t=1}^{n-2}\(\(H_{i\bt_t}\)^\textrm{T}\)^{-1}\cdot J_t(H_{j\bt_t})^\textrm{T}=0, \labell{eq-22}
\ee
where the dummy variable $t'$ has been replaced by $t$. We can continue to transform it into a convenient form for
further attempts to prove, by isolating its real matrix content. Let's define
\be
\Sg_{j\bt_t}\equiv\Sg^{(j)}(t,n-1,\bt_t,n),~~W_{ij}\equiv\frac{1}{\sqrt{\det'(\Phi)(\sg^{(i)})}}\de_{ij},
\ee
then it is clear that $H_{i\bt_t}=W_{ij}\Sg_{j\bt_t}$.
While $W_{ij}$ is a trivial diagonal matrix, $\Sg_{j\bt_t}$ encodes the real matrix content.
Now we can write the LHS of \eqref{eq-22} as
\be
\bal
&\sum_{t=1}^{n-2}W^{-1}\(\(\Sg_{i\bt_t}\)^\textrm{T}\)^{-1}\cdot J_t\((\Sg_{j\bt_t})^\textrm{T}W\)\\
=&\sum_{t=1}^{n-2}W^{-1}\(\(\Sg_{i\bt_t}\)^\textrm{T}\)^{-1}\cdot J_t(\Sg_{j\bt_t})^\textrm{T}\cdot W
+\sum_{t=1}^{n-2}W^{-1}\cdot J_tW\\
=&\sum_{t=1}^{n-2}\(\(\Sg_{i\bt_t}\)^\textrm{T}\)^{-1}\cdot J_t(\Sg_{j\bt_t})^\textrm{T}
+W^{-1}\cdot\(\sum_{t=1}^{n-2}J_t\)W\\
=&\sum_{t=1}^{n-2}\(\(\Sg_{i\bt_t}\)^\textrm{T}\)^{-1}\cdot J_t(\Sg_{j\bt_t})^\textrm{T},
\eal
\ee
where in the third line, the second term vanishes again due to angular momentum conservation. Finally, we are left with
\be
\sum_{t=1}^{n-2}\(\(\Sg_{i\bt_t}\)^\textrm{T}\)^{-1}\cdot J_t(\Sg_{j\bt_t})^\textrm{T}=0,
\ee
which can no longer be further simplified.

To get some sense of this very nontrivial identity, it is helpful to see
the first nontrivial case $n=4$, which corresponds to the first nonempty $\bt_t$. Recall that
\be
\Sg^{(i)}(\ap)=\frac{1}{\sg^{(i)}_{\ap(1),\ap(2)}\ldots\sg^{(i)}_{\ap(n-1),\ap(n)}\sg^{(i)}_{\ap(n),\ap(1)}},
\ee
we have
\be
\bal
&\(\Sg^{(i)}(1,3,2,4)\)^{-1}J_1\Sg^{(j)}(1,3,2,4)+\(\Sg^{(i)}(2,3,1,4)\)^{-1}J_2\Sg^{(j)}(2,3,1,4)\\
=\,&\sg^{(i)}_{13}\sg^{(i)}_{32}\sg^{(i)}_{24}\sg^{(i)}_{41}
J_1\(\frac{1}{\sg^{(j)}_{13}\sg^{(j)}_{32}\sg^{(j)}_{24}\sg^{(j)}_{41}}\)
+\sg^{(i)}_{23}\sg^{(i)}_{31}\sg^{(i)}_{14}\sg^{(i)}_{42}
J_2\(\frac{1}{\sg^{(j)}_{23}\sg^{(j)}_{31}\sg^{(j)}_{14}\sg^{(j)}_{42}}\)\\
=\,&\sg^{(i)}_{13}\sg^{(i)}_{32}\sg^{(i)}_{24}\sg^{(i)}_{41}
(J_1+J_2)\(\frac{1}{\sg^{(j)}_{13}\sg^{(j)}_{32}\sg^{(j)}_{24}\sg^{(j)}_{41}}\)=0,
\eal
\ee
which trivially holds by the antisymmetry of $\sg_{ab}$! But as $n$ increases, even for $n=5$ this identity will
be much more entangled and simple antisymmetry is insufficient for its proof. The potential toolkit for
this purpose includes: (1) relations of spinor derivatives on scattering equations; (2) KK and BCJ relations
of $\Sg_{j\bt_t}$; (3) induction, which may involve contour integration.
We will come back to this point in the future after better understanding the scattering equations
and their solutions.

A last comment is that in \eqref{eq-20}, the anti-holomorphic angular momentum operator $J_{t',\dot{\ap}\dot{\bt}}$
should be generalized to $J_{t',\mu\nu}$ in arbitrary dimensions. Since in 4-dimension
$J_{\mu\nu}\sim\e_{\ap\bt}J_{\dot{\ap}\dot{\bt}}+\e_{\dot{\ap}\dot{\bt}}J_{\ap\bt}$, and the soft theorem must hold
for both holomorphic and anti-holomorphic soft limits, it is more natural to use $J_{\mu\nu}$ as all other quantities
are already defined for arbitrary dimensions.

\section*{Acknowledgments}

The authors would like to thank Qingjun Jin for reading the manuscript.
This work is supported by\\
Qiu-Shi Funding and Chinese NSF funding under contracts No.11135006, No.11125523 and No.11575156.


\end{document}